\documentclass[superscriptaddress,twocolumn,showpacs,amsmath,amssymb]{revtex4}

\usepackage[dvips]{graphicx}
\usepackage{graphicx}

\renewcommand{\vec}[1]{\mbox{\boldmath $#1$}}

\begin{document}

\title{
Non-collective excitations in low-energy heavy-ion reactions:
applicability of the random-matrix model }

\author{S. Yusa}
\affiliation{
Department of Physics, Tohoku University, Sendai 980-8578,  Japan} 

\author{K. Hagino}
\affiliation{
Department of Physics, Tohoku University, Sendai 980-8578,  Japan} 

\author{N. Rowley}
\affiliation{
Institut de Physique Nucl\'{e}aire, UMR 8608, CNRS-IN2P3 et Universit\'{e}
de Paris Sud, 91406 Orsay Cedex, France}

\begin{abstract}
We investigate the applicability of a random-matrix model to the
description of non-collective excitations in heavy-ion
reactions around the Coulomb barrier.
To this end, we study fusion in the reaction $^{16}$O + $^{208}$Pb, 
taking account of the known non-collective excitations in 
the $^{208}$Pb nucleus. 
We show that the random-matrix model for the corresponding
couplings reproduces reasonably well the exact 
calculations, obtained using empirical deformation parameters. 
This implies that the model may provide a powerful method 
for systems in which the non-collective couplings are not so well known. 
\end{abstract}


\maketitle

\section{Introduction}
Heavy-ion reactions around the Coulomb barrier 
often show a behavior that cannot be accounted for by a simple
potential model~\cite{dasgupta,BT98,HT12}. They 
have thus provided a
good opportunity to investigate the role of internal degrees of freedom
in the reaction process.  
One of the well known examples is a large 
enhancement of sub-barrier fusion
cross sections due to the couplings between the relative motion of 
the projectile and target nuclei and their internal degrees of freedom, 
such as surface vibrations for spherical nuclei, 
or rotational motion for nuclei possessing a static, intrinsic deformation. 
It is well recognized that these couplings lead to 
a distribution of potential barriers~\cite{DLW83}, and
a method was proposed by Rowley, Satchler and Stelson to extract
the barrier distributions directly
from experimental fusion cross sections~\cite{RSS91}.
The barrier distributions extracted in this way are 
found to be sensitive to details of the
couplings, often showing a characteristic structured 
behavior~\cite{dasgupta,L95,LRL93}.
Similar heavy-ion barrier distributions can also be defined for 
large-angle quasi-elastic scattering~\cite{timmers,HR04}.

In order to 
analyze experimental data for these
low-energy heavy-ion reactions, 
the coupled-channels method has been employed 
as a standard approach~\cite{ccfull,HT12}. 
This method describes the reaction in terms of 
the internal excitations of the colliding nuclei, 
representing the total wave function of the system
as a superposition of wave functions for the relevant reaction channels.
Conventionally, a few low-lying collective excitations, that are 
strongly coupled to the ground state, 
are taken into account in these calculations.
Such analyses have successfully accounted for
the strong enhancement of sub-barrier fusion cross sections, and have
successfully reproduced the structure of the fusion and quasi-elastic 
barrier distributions for many systems~\cite{dasgupta,HT12}. 

Recently, quasi-elastic barrier distributions have been 
measured for the $^{20}$Ne +$^{90,92}$Zr systems~\cite{piasecki}. 
The corresponding coupled-channels calculations 
show that the main structure of these barrier 
distributions is determined by the rotational excitations of the 
strongly deformed nucleus $^{20}$Ne. The calculated barrier distributions 
are in fact almost identical for the two systems, even when 
the collective excitations of the Zr isotopes are taken into account. 
It was, therefore, surprising when the 
two experimental barrier distributions were found to be
different in an important respect. 
That is, the barrier distribution
for $^{20}$Ne + $^{92}$Zr exhibits a much more smeared behavior
than that for the $^{20}$Ne + $^{92}$Zr system. 
The origin of this difference has been 
conjectured in Ref.~\cite{piasecki} 
to be the multitude of non-collective excitations of the Zr isotopes, 
that are generally ignored in a coupled-channels analysis.
In fact, the two extra neutrons in 
the $^{92}$Zr nucleus lead to 
a significantly larger number of non-collective excited states 
compared with $^{90}$Zr, since this latter possesses an $N=50$ closed shell 
(the difference is reflected by the number of known states up to an excitation 
energy of 5~MeV; one finds 35 for $^{90}$Zr and 87 for $^{92}$Zr~\cite{bnl}). 

There are many ways to describe 
non-collective excitations in heavy-ion reactions 
\cite{YHR10,YHR12,DHD08, B10, B102,KPW76,AKW78,BSW78,akw1,akw2,akw3,akw4}. 
In the 1970's, Weidenm\"uller {\it et al.} introduced  
a random-matrix model for such excitations in order 
to study deep inelastic 
collisions~\cite{KPW76,AKW78,BSW78,akw1,akw2,akw3,akw4}. 
In Ref.~\cite{YHR10}, we have used a similar model in a schematic 
one-dimensional
barrier-penetration problem, in order to study the role of  
these non-collective excitations in low-energy reactions. 
On the other hand, in Ref.~\cite{YHR12}, 
we have explicitly taken into account 
in the coupled-channels formalism 
all of the 70 known non-collective states 
in $^{208}$Pb below 7.382~MeV~\cite{WCHM75,LBF73} without resorting 
to the random-matrix model, 
and have analysed in this way the experimental data for the 
$^{16}$O+$^{208}$Pb reaction. Although some discrepancies 
between the experimental and
theoretical barrier distributions remain after the inclusion of 
non-collective excitations, we have shown in Ref.~\cite{YHR12} that
these excitations play a more important role as the incident
energy increases. We have also compared there the role of non-collective
excitations in the fusion and quasi-elastic barrier distributions,
and have shown that they affect both distributions in 
a similar fashion.

Given that exact calculations with a realistic spectrum for 
non-collective states is possible here, it is intriguing 
to also apply the random-matrix model to this system 
in order to test its applicability. 
This theoretical test is the main aim of this paper, and we achieve it 
by comparing our new results with those obtained in Ref.~\cite{YHR12}. 
Note that, in contrast to $^{208}$Pb, the properties of 
the non-collective states in $^{90,92}$Zr 
are not known sufficiently well. As we will show in this paper, the
random-matrix model 
provides a good method for a description of non-collective excitations 
in such a situation. 

The paper is organized as follows. 
In Sec. II, we explain the coupled-channels 
formalism with non-collective excitations based on the random-matrix model. 
In Sec.~III, we discuss the strength distribution and 
fusion cross sections obtained with this model. 
We then compare these with calculations using the more exact 
couplings and discuss the applicability of the random-matrix model. 
The paper is summarised in Sec. IV.

\section{Coupled-channels method with non-collective excitations}

In order to describe 
internal excitations during the reaction process, 
we assume the following Hamiltonian, 
\begin{eqnarray}
  H = -\frac{\hbar^2}{2\mu}\nabla^2 + V_{\rm rel}(r) + H_0(\{\xi\}) 
+ V_{\rm coup}(\vec{r},\{\xi\}), 
\end{eqnarray}
where $\vec{r}$ is the separation of the 
projectile and target nuclei, and $\mu$ is the reduced mass. 
In this equation, $H_0(\{\xi\})$ is the intrinsic Hamiltonian
with $\{\xi\}$ representing a set of internal degrees of freedom. 
The optical potential for the relative motion is $V_{\rm rel}(r)$, 
and it includes an imaginary part to simulate the fusion process 
(that is, strong absorption into compound-nucleus degrees of 
freedom inside the Coulomb barrier).
The coupling Hamiltonian between 
the relative motion and the intrinsic degrees of freedom 
is denoted by $V_{\rm coup}(\vec{r},\{\xi\})$.  

The coupled-channels equations for
this Hamiltonian are obtained by expanding the 
total wave function in terms of the eigenfunctions of $H_0(\{\xi\})$. 
The equations read, 
\begin{eqnarray} 
 \left[-\frac{\hbar^2}{2\mu}\frac{d^2}{dr^2} +  \frac{J(J+1)\hbar^2}{2\mu r^2}
+ V_{\rm rel}(r) + \epsilon_n - E\right]u_n^{J}(r) \nonumber&&\\
 +  \sum_m V_{nm}(r)u_m^{J}(r) = 0,
\end{eqnarray}
where $\epsilon_n$ is the excitation energy for the $n$-th channel.
In deriving these equations, we have employed 
the iso-centrifugal approximation
\cite{HT12,LR84,NRL86,NBT86,ELP87,T87,GAN94}.
In this approximation, 
the orbital angular momentum in the centrifugal potential 
is replaced by the total angular momentum~$J$, thereby considerably 
reducing the dimension of the coupled-channels equations. 

In solving these equations, 
we impose the following asymptotic boundary
conditions,
\begin{eqnarray}
  u_n^{J}(r) \rightarrow H_{J}^{(-)}(k_n r)\delta_{n,0} -
  \sqrt{\frac{k_0}{k_n}}S_n^{J}H_{J}^{(+)}(k_n r), 
\end{eqnarray}
for $r \rightarrow \infty$, together with the regular boundary 
condition at the origin.
Here, $k_n = \sqrt{2\mu(E - \epsilon_n)/\hbar^2}$ is the wave number for the
$n$-th channel, where $n=0$ 
corresponds to the entrance channel. 
$S_n^{J}$ is the nuclear $S$-matrix, 
and $H_{J}^{(-)}(kr)$ and $H_{J}^{(+)}(kr)$ are the incoming and
the outgoing Coulomb wave functions, respectively. 
The fusion cross sections are then obtained as 
\begin{eqnarray}
\sigma_{\rm fus}(E) = \frac{\pi}{k_0^2}\sum_J(2J+1)
\left(1 - \sum_n\left|S_{n}^{J}\right|^2\right).
\end{eqnarray}

In the random-matrix model~\cite{KPW76,AKW78,BSW78,akw1,akw2,akw3,akw4}, 
one assumes an ensemble of coupling matrix
elements whose first moment satisfies 
\begin{align}
\overline{V_{nn^\prime}^{II^\prime}(r)} = 0,
\end{align}
while the second moment satisfies 
\begin{align}
&\ \ \ \ \ \overline{V_{nn^\prime}^{II^\prime}(r)V_{n^{\prime\prime}n^{\prime\prime\prime}}^{I^{\prime\prime}I^{\prime\prime\prime}}(r^\prime)} \nonumber \\
&=  \left\{\delta_{nn^{\prime\prime}}\delta_{n^{\prime}n^{\prime\prime\prime}} 
          \delta_{II^{\prime\prime}}\delta_{I^{\prime}I^{\prime\prime\prime}}
   + \delta_{nn^{\prime\prime\prime}}\delta_{n^{\prime}{n^{\prime\prime}}}
                      \delta_{II^{\prime\prime\prime}}\delta_{I^{\prime}{I^{\prime\prime}}}
   \right\} \nonumber
  \\ &\ \ \ \ \times  \sqrt{(2I+1)(2I^\prime+1)}\,  \nonumber
  \sum_{\lambda}
   \left(
     \begin{array}{ccc}
       I & \lambda & I^{\prime} \\
       0 & 0 & 0
     \end{array}
   \right)^{2} \nonumber\\
&\ \ \ \ \times
\alpha_{\lambda}(n,n^\prime;I,I^\prime;r, r^\prime).
\label{second_moment}
\end{align}
Here, $I$ is the spin of the intrinsic state labeled by $n$, 
and $\alpha_{\lambda}$ is the coupling form factor.

In this paper, for simplicitly, we assume that 
the non-collective excitations couple only to the ground state, 
as in the linear coupling approximation employed in our 
previous work~\cite{YHR12}. 
For the form factor $\alpha_{\lambda}$, 
we assume the following dependence
\begin{eqnarray}
\alpha_{\lambda}(n,0;I,0;r, r^\prime) =
\frac{w_\lambda}{\sqrt{\rho(\epsilon_n)}}
e^{-\frac{\epsilon_n^2}{2\Delta^2}}
e^{-\frac{(r-r^{\prime})^2}{2\sigma^2}}
h(r)h(r^{\prime}),
\label{form_factor}
\end{eqnarray}
where $\rho(\epsilon_n)$ is the level density at an excitation energy
$\epsilon_n$, and ($w_{\lambda}, \Delta, \sigma$) are adjustable parameters.
The appearance of the level density in the denominator
reflects the complexity of the
non-collective excited states, as discussed in Ref.~\cite{akw1}.
For the function $h(r)$,
we adopt the derivative of the Woods-Saxon potential, that is,
\begin{eqnarray}
h(r) = \frac{e^{(r-R)/a}}
{\left[1 + e^{(r-R)/a}\right]^2}.
\end{eqnarray}
Note that this choice of the form factor
corresponds to the coupling Hamiltonian in the 
linear coupling approximation derived from the
Woods-Saxon potential.

\section{Applicability of random-matrix model}

\subsection{Strength distribution}

\begin{figure}[t]
  \center
  \begin{minipage}[t]{78mm}
    \includegraphics[clip,keepaspectratio,width=78mm]{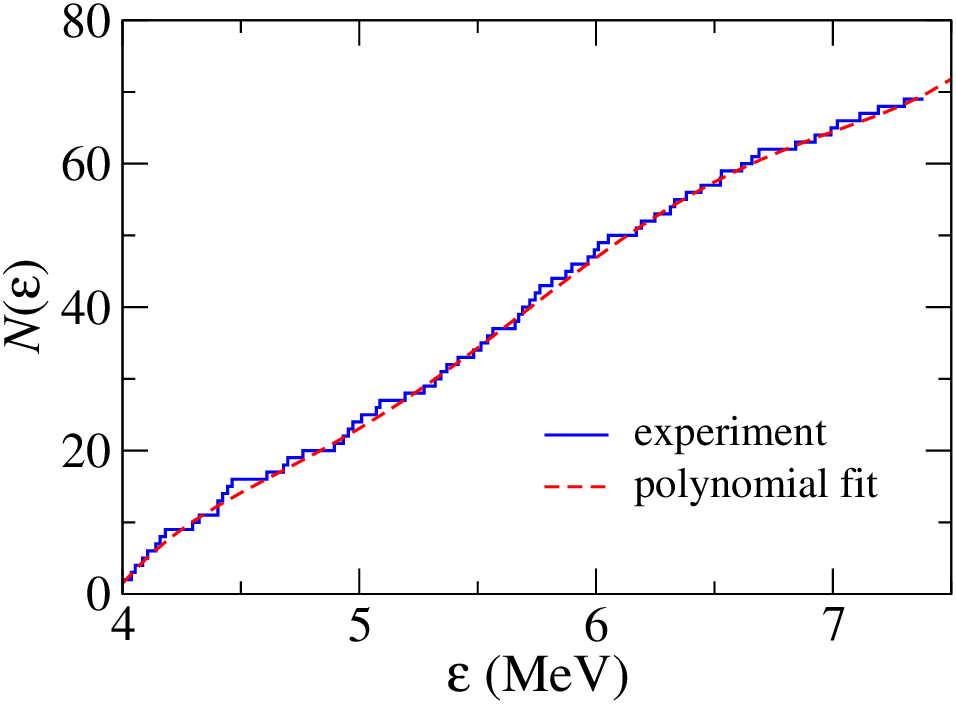}
    \caption{(Color online) The number of levels of the $^{208}$Pb nucleus 
    up to the excitation energy $\epsilon$
    as a function of $\epsilon$. The histogram
    represents the experimental data~\cite{WCHM75}, while the dashed 
line shows its fit with a polynomial function up to the sixth order. } 
    \label{fig7.5}
  \end{minipage}
  \hspace{0.5cm}
  \begin{minipage}[t]{78mm}
    \includegraphics[clip,keepaspectratio,width=78mm]{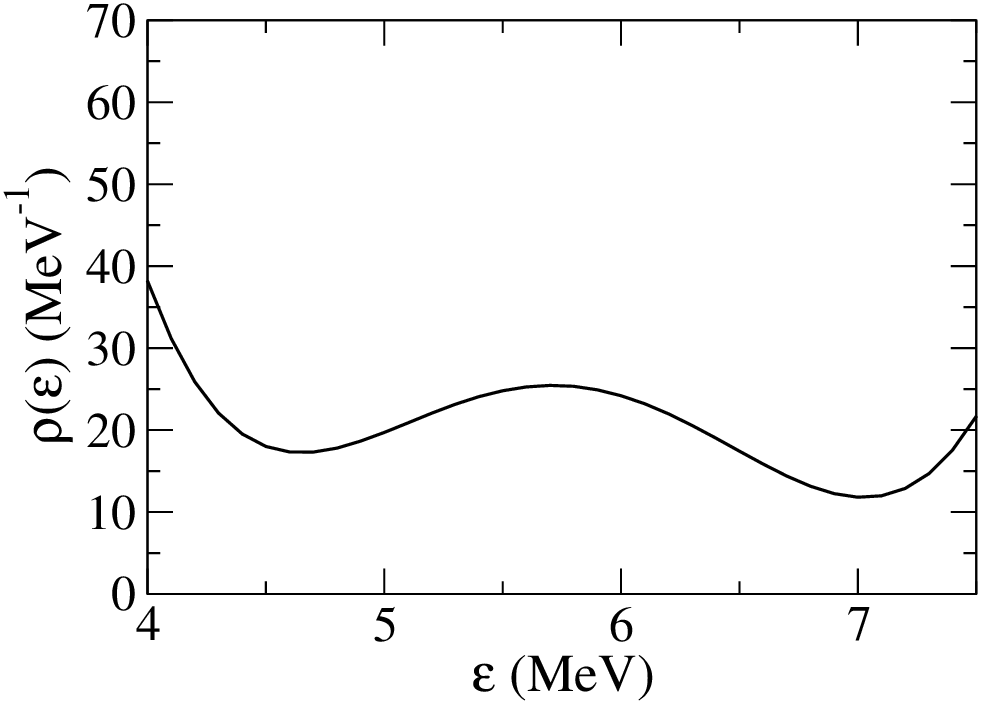}
    \caption{The continuous level density for $^{208}$Pb 
obtained as a first derivative
             of the fitting function $f(\epsilon)$
             shown in Fig.~\ref{fig7.5}.}
    \label{fig7.6}
  \end{minipage}
\end{figure}

Let us now apply the random-matrix model 
to the $^{16}$O+$^{208}$Pb reaction and discuss its 
applicability. We first discuss the strength distribution for the 
non-collective excitations in $^{208}$Pb obtained with the random-matrix model. 
To this end, we define the strength distribution as 
\begin{align}
b_I &= 
\sqrt{
\sum_{\lambda}
  \left(
    \begin{array}{ccc}
      0 & \lambda & I \\
      0 & 0 & 0
    \end{array}
  \right)^2
  \sqrt{\frac{2I+1}{\rho(\epsilon)}}
  e^{-\frac{\epsilon^2}{2\Delta^2}} 
  } \nonumber \\
&=
\sqrt{
  \sqrt{\frac{2I+1}{\rho(\epsilon)}}
  e^{-\frac{\epsilon^2}{2\Delta^2}}
}.
\label{strength_dist}
\end{align}
This quantity essentially corresponds to the 
\lq\lq square root\rq\rq\, of Eq.~(\ref{second_moment}),
except for an overall scale factor 
(here we have assumed $w_{\lambda} = w$ for all $\lambda$ and omitted 
$w_{\lambda}$ in the definition for the strength function).

The level density in Eq.~(\ref{strength_dist}) is 
treated in the following way. 
It is originally 
defined by 
\begin{eqnarray}
\rho(\epsilon) = \sum_n \delta(\epsilon - \epsilon_n)
\end{eqnarray}
for a discrete spectrum.
For practical purposes, 
we define the function
\begin{eqnarray}
N(\epsilon) =  \int^\epsilon_0 \rho(\epsilon^\prime)d\epsilon^\prime 
=\sum_n\theta(\epsilon-\epsilon_n),
\label{num_levels}
\end{eqnarray}
that gives the number of levels up to the excitation energy
$\epsilon$. We fit this function with a polynomial in $\epsilon$, 
and then define a continuous level density by differentiating 
this polynomial. 
Figure \ref{fig7.5} shows the experimental $N(\epsilon)$ for $^{208}$Pb 
\cite{WCHM75} 
in the interval between 4~MeV and 7.5~MeV (solid line) and its fit 
with a polynomial
$\displaystyle f(\epsilon) = \sum_{n=0}^6 a_n\epsilon^n$ (dashed line). 
The values of $a_n$ are
$a_0=-7479, a_1=6969$ (MeV$^{-1}$), $a_2=-2612$ (MeV$^{-2}$),
$a_3=497.5$ (MeV$^{-3}$), $a_4=-49.59$ (MeV$^{-4}$),
$a_5=2.347$ (MeV$^{-5}$), and $a_6=-0.03632$ (MeV$^{-6}$).
The continuous level density, 
$\rho(\epsilon)=df(\epsilon)/d\epsilon$, is shown in 
Fig.~\ref{fig7.6}.

\begin{figure}[t]
\center
  \includegraphics[clip,keepaspectratio,width=80mm]{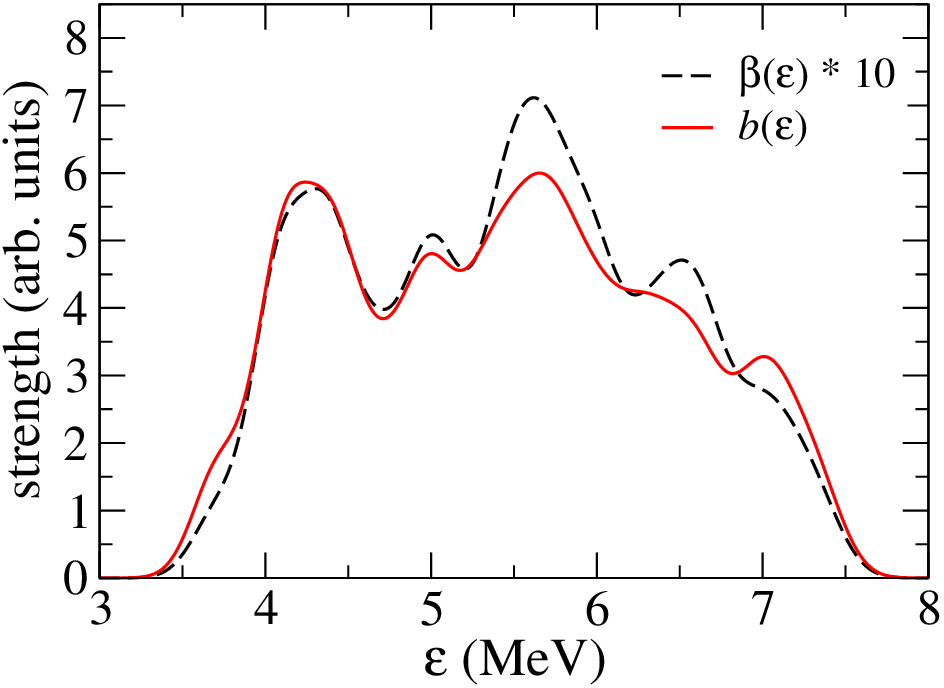}
  \caption{(Color online) The 
strength distributions for $^{208}$Pb 
as a function of excitation energy $\epsilon$.
The dashed line shows the distribution of the experimental 
deformation parameters, while the solid line
is obtained based on the random-matrix model using 
Eq.~(\ref{strength_dist}).
Both distributions are smeared with a Gaussian function with a width 
of 0.15~MeV. An overall scaling factor is introduced to the dashed 
line as the dimension is different between the two curves (see text).}
  \label{comp_str_dist}
\end{figure}

The strength distribution $b_{I}$ 
calculated with this level density
is shown in Fig.~\ref{comp_str_dist} by the solid line 
as a function of excitation energy $\epsilon$. 
The parameter $\Delta$ in Eq.~(\ref{strength_dist}) is chosen to
be 7~MeV, as in Refs.~\cite{akw3,akw4}.
For comparison, the figure also shows the 
distribution of the experimental deformation parameters $\beta_{I}$ 
\cite{WCHM75}, smeared with a gaussian function with a width of 
0.15~MeV (dashed line). We have also performed the same smearing 
for the strength distribution $b_I$. 
Also, since the dimensions of $\beta_{I}$ and $b_{I}$ are not the same, 
the deformation parameters $\beta_I$ are scaled by a factor 10 
so that the heights of the first peaks at about 4.3~MeV match 
one another. 
Although there exists a small deviation 
for the peaks between 5~MeV and 7~MeV, 
the overall structure of the
strength distribution is well reproduced by this model. 

\subsection{Fusion cross sections}

\begin{figure}[t]
\center
  \includegraphics[clip,keepaspectratio,width=85mm]{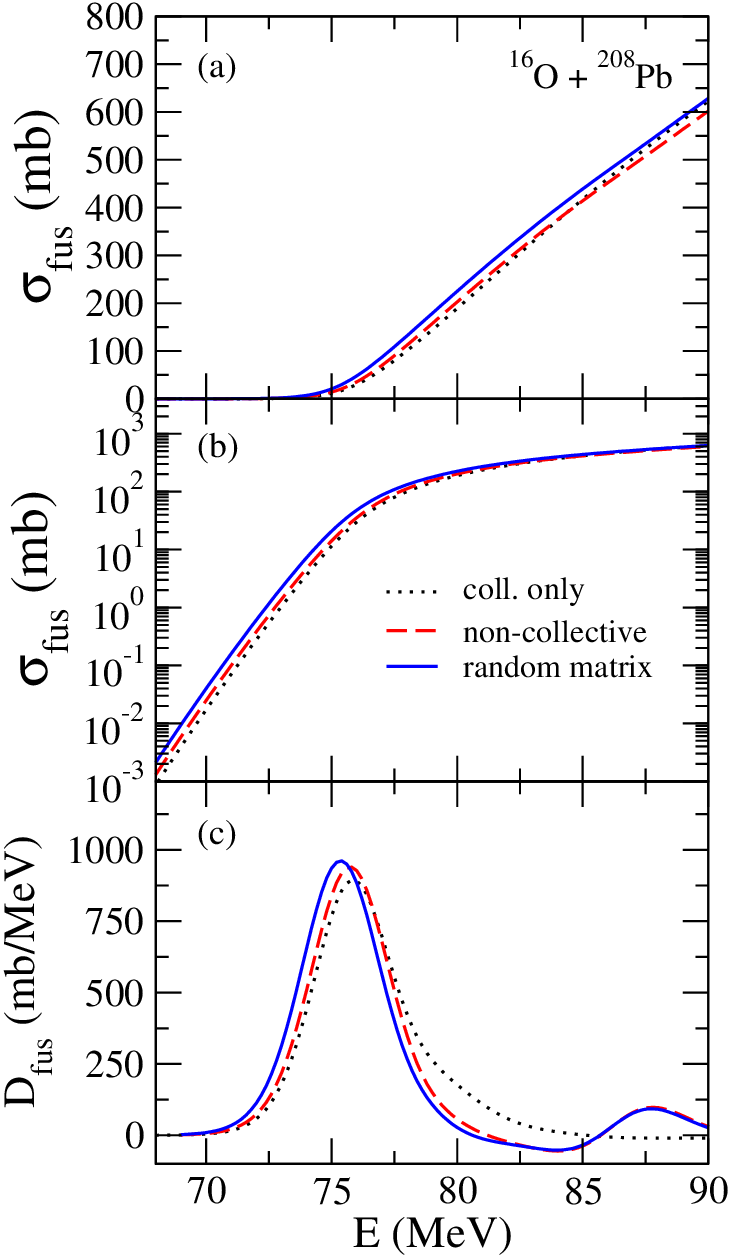}
  \caption{(Color online) (a) and (b) Fusion cross sections $\sigma_{\rm fus}$ 
and (c) fusion barrier distributions 
$D_{\rm fus}(E)=d^2{E\sigma_{\rm fus}(E)}{dE^2}$ 
for the $^{16}$O + $^{208}$Pb system obtained from three different calculations.  
Dashed lines show results obtained with the experimental, non-collective
deformation parameters, whereas the solid lines are obtained 
from the random-matrix model. Dotted lines result from calculations 
that include only the $^{208}$Pb collective excitations. }
 \label{fusion}
\end{figure}

The strength distribution discussed in the previous subsection 
determines the coupling strength to each excited state. 
Let us then examine how the random-matrix model 
can be compared with the exact results 
in terms of the fusion cross sections for the $^{16}$O + $^{208}$Pb system.
For this purpose, 
we use the same Woods-Saxon potential for the nuclear potential as 
in Ref.~\cite{YHR12}; it has 
a surface diffuseness $a=0.671$~fm, a radius $R = 8.39$ fm and
a depth $V_0$ = 550~MeV. 
For the couplings to the collective excitations, we take
into account the vibrational $3^-$ state at 2.615~MeV,
the $5^-$ state at 3.198~MeV, and the $2^+$ state
at 4.085~MeV in $^{208}$Pb.
The octupole mode is included up to the two-phonon states, while 
the other, weaker, vibrational modes are taken into account only up to 
their one-phonon states.
The deformation parameters for these vibrational modes are
estimated from the measured electromagnetic transition
probabilities. They are
$\beta_3 $ = 0.122, 
$\beta_5 $ = 0.058, and
$\beta_2 $ = 0.058
together with a radius parameter 
of $r_0$=1.2~fm.
Although we took into account the octupole phonon state of
$^{16}$O in our preivous study~\cite{YHR12}, for simplicity 
we do not include it in the present calculations, since its
effect can be well described by an adiabatic
renormalization of the potential depth~\cite{HT12,THAB94}.
For the parameter $\sigma$ 
in Eq.~(\ref{form_factor}), 
we follow Refs.~\cite{akw3, akw4} and use $\sigma = 4$ fm. 
On the other hand, the parameter
$w_{\lambda}=w$ is chosen to be 
$w = 38000$~MeV$^{3/2}$ 
so that the height of the main peak in the 
fusion barrier distribution
is reproduced by the random-matrix model.  

Figures \ref{fusion} (a) and \ref{fusion} (b) show 
the $^{16}$O+$^{208}$Pb fusion excitation function 
on linear and logarithmic scales respectively. 
The dashed lines show the results obtained with 
the {\em measured} deformation parameters for 
the non-collective excitations, while the solid lines show 
the results obtained using the random-matrix approximation.
For comparison, the dotted lines show
results that account only for the collective excitations. 
Although a small overall shift can be seen, it is clear that 
the random-matrix model 
reproduces the exact results reasonably well. 

In order to highlight the energy dependence, 
Fig.~\ref{fusion}~(c) shows the fusion barrier distribution
$D_{\rm fus}(E)=d^2(E\sigma_{\rm fus})/dE^2$ 
\cite{dasgupta,HT12,RSS91,L95}. 
Although the main peak is slightly shifted in energy, 
this confirms that the random-matrix model 
reproduces well the exact results. 
That is, with respect to the dotted line, the change in the energy 
dependence of fusion cross sections due to the non-collective excitations 
is similar in the two calculations. In particular, 
both barrier distributions are smeared out in a similar way 
at energies around 80~MeV, and both calculations yield a similar 
second peak around 87.5~MeV. (We note that if the strength $w_0$
was somewhat larger, the second peak could appear at even
higher energies, possibly reflecting the broad bump seen at around 97~MeV 
in the experimental data.) 

As we have argued in Ref.~\cite{YHR10}, the
higher-energy peaks in the barrier distribution
are affected more by non-collective
excitations than are the lower-energy peaks. Unfortunately this is not
easy to see in Fig.~\ref{fusion} because the peaks obtained with purely 
collective couplings are not resolved.
This difference can, however, be easily understood using perturbation theory. 
That is, the eigenchannels corresponding to the higher-energy peaks in the
barrier distribution couple more strongly to the non-collective states
via their ground state component simply because the energy 
differences are smaller. Higher peaks are thus redistributed more,
effectively removing much of their strength from that region of energy.

From these calculations, it is evident that 
the effects of non-collective excitations are not 
sensitive to details of the non-collective couplings, and 
that the random-matrix model is applicable to the 
description of non-collective excitations, 
so long as the relevant parameters are chosen appropriately. 

\section{Summary}

We have investigated the applicability of the 
random-matrix model for the description of non-collective
excitations in low-energy heavy-ion reactions.
To this end, we have calculated the fusion excitation function for the 
$^{16}$O  +$^{208}$Pb system, 
where the role of the 
non-collective excitations has already been investigated
in our previous study using empirical deformation parameters. 

We have first shown that the coupling strength distribution
obtained with the random-matrix model agrees well with the
experimental distribution. 
The fusion cross section and barrier 
distribution for the $^{16}$O + $^{208}$Pb system obtained 
with empirical non-collective couplings are also 
well reproduced by the random-matrix model with 
appropriately chosen parameters. 
These results provide a validation of 
the random-matrix model for the description of
non-collective couplings.

For the $^{208}$Pb nucleus, detailed properties of non-collective 
states are known over a large energy range. 
However, this is not always the case for other systems.
That is, for many nuclei, even though the energies and
spin-parity may be relatively well known for many non-collective states, 
the coupling strengths are poorly determined. 
In such a situation, the present study suggests that
the random matrix model provides a powerful tool to treat
these coupling strengths.
A good example is the quasi-elastic barrier distribution for 
the $^{20}$Ne + $^{90,92}$Zr systems, where it has been suggested 
that non-collective excitations may play an important role. 
Analyses for these systems within the random-matrix model 
are under way. We shall report the results 
in a separate publication~\cite{YHR13}.

\begin{acknowledgments}
This work was supported by the Global COE Program
``Weaving Science Web beyond Particle-Matter Hierarchy'' at
Tohoku University,
and by the Japanese
Ministry of Education, Culture, Sports, Science and Technology
by Grant-in-Aid for Scientific Research under
the program number (C) 22540262.
\end{acknowledgments}

\end{document}